# Birefringence analysis of multilayer leaky cladding optical fibre

**L Labonté[1], V Rastogi[2●], A Kumar[2], B Dussardier[1] and G Monnom[1]**

[1]Laboratoire de Physique de la Matière Condensée
 Université de Nice-Sophia Antipolis, Parc Valrose 06108 Nice, France
[2]Department of Physics, Indian Institute of Technology, Roorkee 247 667, India

E-mail: vipul.rastogi@osamember.org

**Abstract.** We analyze a multilayer leaky cladding (MLC) fibre using the finite element method and study the effect of the MLC on the bending loss and birefringence of two types of structures: i) a circular-core large-mode area structure and ii) an elliptical-small-core structure. In a large-mode-area structure, we verify that the multi-layer leaky cladding strongly discriminates against higher order modes to achieve single-mode operation, the fibre shows negligible birefringence, and the bending loss of the fibre is low for bending radii larger than 10 cm. In the elliptical-small-core structure we show that the MLC reduces the birefringence of the fibre. This prevents the structure from becoming birefringent in case of any departures from circular geometry. The study should be useful in the designs of MLC fibres for various applications including high-power amplifiers, gain flattening of fibre amplifiers and dispersion compensation.

## 1. Introduction

Leaky modes in fibres have always been a matter of interest in designing photonic devices such as bent waveguides [1], depressed inner cladding (DIC) single-mode (SM) fibres [2] and polarizers [3]. Recently leaky structures have shown their potential for designing fibres for large-mode area (LMA) SM operation and gain equalization of optical fibre amplifiers. We have recently designed LMA SM fibres for high power delivery by tailoring the cladding refractive index profile in radial or angular directions [4-7]. All the modes of such fibres suffer from finite leakage loss. A high differential leakage loss between the fundamental and higher order modes and a nominal loss to the fundamental mode ensure the effective single mode operation in these designs. In one such design, namely the Segmented Cladding Fibre (SCF), a periodic variation of low- and high-index segments in the angular direction inside the cladding region has been used to achieve the LMA operation [4]. However, fabrication of SCF in silica could still not be possible by modified chemical vapour deposition (MCVD) technique. Therefore few simpler leaky MCVD LMA designs have been proposed: a graded-index cladding with radially rising refractive index [5], a cladding made of periodically arranged low-index trenches of varying strength in an otherwise high-index medium [6], and a dual core leaky design based on resonance coupling effect [7]. Some small-core resonant multilayer leaky cladding (MLC) designs have also been proposed and have shown their potential applications in gain equalization of erbium-doped fibre amplifiers [7] and in dispersion compensators [8]. An MLC fibre with 25 micron diameter and 0.11 numerical aperture (NA) has also been realized at the University of Nice, France, by the MCVD technique [9]. Until now these designs have been studied using numerical techniques, such as the transfer matrix method (TMM) [10,11] and the radial effective-index method (REIM) [4,12]. The TMM and REIM are fast and easy to implement but do not calculate the bending loss and birefringence of the structure, which might be important in LMA designs [6,7] and small-core resonant designs [7,8],

---

[●]Author to whom any correspondence should be addressed.

respectively. In the present work, we carry out the analysis of the MLC fibre using the finite element method (FEM), which is considered as one of the most powerful and versatile methods in many branches of engineering to solve complicated problems. General equations can be easily solved in an approximate manner by using FEM as numerical tool. In the finite element approach, the problem domain can be suitably divided into many triangles of different shapes and sizes. This makes the FEM preferable when compared with the finite difference method (FDM) which not only uses inefficient regular spaced meshing but also cannot represent curved dielectric interfaces adequately. Introduction of perfectly matched layer (PML) to the FEM can also enable the calculation of leakage loss of the fibre. In this paper we have carried out FEM analysis of an MLC fibre such as that proposed in [6] and have compared our results with those obtained by the TMM. An excellent agreement on effective indices and the leakage losses has been obtained between both methods. We have also studied the bending loss of the fibre by employing PML and have found that the fibre suffers from small bending loss (0.006 dB/m) at the operating wavelength (1.55 μm) for the bending radius 10 cm. The analysis has also been extended to a small-elliptical-core MLC fibre and the influence of the MLC on the birefringence and bending loss of the fibre has been investigated. We have shown that the introduction of the MLC to an otherwise standard elliptical-core fibre can reduce the birefringence. The bending loss of such a fibre is sensitive to polarization and the sensitivity can be controlled by the cladding profile. The study should be useful in the designs of MLC fibres for high power optical amplifiers, high power lasers, gain-equalization of optical amplifiers and dispersion compensation.

## 2. Finite element analysis

An MLC fibre consists of a uniform guiding core of refractive index $n_1$ and radius $a$. The cladding of the fibre is characterized by periodically spaced low- and high-index layers. Periodicity of low-index (depressed) layers of width $d$ is denoted by $\Lambda$. All high-index cladding layers have width $(\Lambda - d)$ and refractive index $n_1$, while the refractive indices of depressed cladding layers follow a power law profile given by

$$n^2(r) = n_1^2 \left[ 1 - 2\Delta \left( \frac{|b-r|}{|b-a|} \right)^q \right] \qquad a < r < b \qquad (1)$$

where $\Delta = \dfrac{n_1^2 - n_2^2}{2n_1^2}$ is the relative index difference between the core and the first cladding layer, $n_2$ being the refractive index of the first depressed layer of the cladding. In case of large core radius, owing to the strong light confinement in the core region, the fibre NA can be defined by $n_1 \sqrt{2\Delta}$. $q$ is the profile parameter, which decides the shape of the power-law profile. The refractive index of the infinitely extended outermost region beyond $r = b$ is assumed to be $n_1$. The transverse cross-section and the refractive index (RI) profile of the MLC fibre are shown in figure 1.

We have used the commercial software COMSOL to carryout FEM analysis of the MLC fibre. The mesh has been chosen so as to satisfy desired accuracy in the calculation of birefringence for the simulated structure. In our finest mesh, the fibre cross-section has been divided into 500000 triangular elements. The maximum element size was λ/7, where λ is the wavelength. We have used a circle of 90 μm radius as the computational window with a typical 10 μm thick PML. An accuracy of $1\times10^{-6}$ in birefringence has been achieved using this mesh. We have calculated the values of leakage losses and effective indices of the first two modes of the MLC fibre having various parameters listed in Table 1.

The values of effective indices calculated by the FEM matched with those calculated by TMM up to 6 significant digits. The leakage losses calculated by both methods have been plotted as a function of profile shape parameter $q$ for $LP_{01}$ and $LP_{11}$ modes as shown in figure 2. We can see a very good agreement between both methods.

### 2.1 Bending loss

Bending, which is unavoidable in most of the practical applications, imposes a significant limitation on the applicability of the fibre. It is therefore important to study the bending performance of the fibre. The study carried out on this fibre in Ref. [6] predicted small bending loss of the fibre owing to its high NA, but it could not be quantified. Here, we have calculated the bending loss of the fibre by using the conformal transformations [13] and employing the PML. Using these transformations the bent fibre can be represented by a corresponding straight fibre with an effective refractive-index distribution given by:

$$n_{eff}^2 = n^2 \left( 1 + 2\frac{r}{R}\cos\phi \right) \tag{2}$$

where $n$, $R$, $r$, $\phi$ are the material refractive index, the bending radius and the first two coordinates in a cylindrical polar co-ordinate system, respectively.

We have carried out calculations of bending loss for the fibre parameters listed in Table 1 and have plotted it as a function of bend radius in figure 3 for profile parameter $q = 1$. It can be seen that for $R = 10$ cm the bending loss is about 0.006 dB/m and settles down for large bending radii to a value of $2.3\times10^{-4}$ dB/m, which corresponds to the value of leakage loss of the straight fibre. The losses of $x$- and $y$- polarized modes are almost the same because the geometry of the core is perfectly circular. The bending loss follows the same trend for higher values of $q$. However, the magnitude of bending loss increases with $q$ as the modal field spreads more into the leaky cladding. For example: bending loss of fibre increases from 0.036 dB/m to 0.5 dB/m as we increase the value of $q$ from 1 to 2 for $R = 7.5$ cm.

## 2.2 Elliptical-core MLC leaky fibre

After analyzing the bending performance of the large-core MLC fibre, we have extended our analysis to small core MLC fibre designs and have studied the effect of a leaky cladding on the geometrical birefringence of the elliptical core fibre and the polarization dependent bending loss. The birefringence of the fibre is defined by $B = n_x - n_y$, where $n_x$ and $n_y$ represent the effective indices of the two orthogonally polarized fundamental modes. To study the effect of MLC on birefringence, we have considered a fibre similar to the one shown in figure 1 but with an elliptical core having semi-major axis $a_x$ and semi minor axis $a_y$. To investigate the effect of MLC on polarization properties of the fibre, we have first calculated the birefringence of a standard 2-layer elliptical core fibre with $n_1 = 1.44439$, $\Delta = 1.6$ %, $a_x = 3.75$ μm, and $a_y = 1.25$ μm. We have then introduced the MLC to the elliptical core and have calculated the birefringence for $d = 3$ μm and 5 μm for a given value of $\Lambda = 9$ μm. The results are shown in figure 4. We can see that the introduction of MLC to an otherwise two-layer step-index fibre can bring down the birefringence to some extent, which can be controlled by the profile parameter.

We have also studied the polarization dependence of the bending loss of the elliptical MLC fibre. Note that the curvature is about the small axis of the elliptical core. We have plotted the bending losses of two orthogonally polarized modes as a function of bending radius $R$ for $q = 1$ as shown in figure 5. We can note that the $x$-polarized mode shows slightly more bending loss. We have also studied the effect of bending on the birefringence. To study this we have chosen a fibre with $q = 5$ in order to have significant magnitude of birefringence and have plotted the variation of the birefringence as a function of $R$ in figure 5. We can see that the birefringence of the fibre is hardly affected by bending for bending radii larger than 6 cm, as the bending of the fibre does not significantly alter the asymmetry of the fibre.

## 3. Conclusion

We have carried out  birefringence analysis of an MLC leaky fibre by the FEM. We have implemented PML to calculate the leakage loss and bending loss of the MLC structure. FEM results agree well with those obtained by the TMM for a circular- core MLC fibre. The study of the bending performance of the large-core MLC fibre reveals that the fibre has small bending loss of 0.006 dB/m for 10 cm bending radius. In the small-elliptical-core leaky fibre the MLC is shown to reduce the birefringence of the fibre. The bending of the fibre beyond radii larger than 6 cm does not significantly affect the birefringence. The study should be useful in designing MLC fibre based components and devices.

### Acknowledgement

Ajeet Kumar acknowledges Council of Science and Industrial Research (CSIR), India for Senior Research Fellowship (SRF). This study was partially supported by an Indo-French "*Projet de Recherche en Réseau*" (P2R) « R&D on Specialty Optical Fibers and Fiber-based Components for Optical Communications » funded by the Ministry of Foreign Affairs, the *Centre National de la Recherche Scientifique* (France) and the Department of Science and Technology (India).

**References**

[1]. Thyagarajan K, Shenoy M R and Ghatak A K 1987 Accurate numerical method for the calculation of bending loss in optical waveguides using a matrix approach *Opt. Lett.* **12** 296-8

[2]. Kawakami S and Nishida S 1974 Characteristics of a doubly clad optical fiber with a low-index inner cladding *IEEE J. Quantum Electron.* **QE-10** 879-87

[3]. Thyagarajan K, Diggavi S and Ghatak A K 1987 Analytical investigation of leaky and absorbing planar structures *Opt. Quantum Electron.* **19** 131-7

[4]. Rastogi V and Chiang K S 2001 Propagation characteristics of a segmented cladding fiber *Opt. Lett.* **26** 491-3

[5]. Rastogi V and Chiang K S 2003 Leaky optical fiber for large mode area single mode operation *Electron. Lett.* **39** 1110-2

[6]. Kumar A and Rastogi V 2008 Design and analysis of a multilayer cladding large-mode area optical fibre *J. Opt. A: Pure Appl. Opt.* **10** 015303

[7]. Kumar A, Rastogi V, Kakkar C and Dussardier B 2008 Co-axial dual-core resonant leaky fibre for optical amplifiers *J. Opt. A: Pure Appl. Opt.* **10** 115306

[8]. Rastogi V and Kumar A 2009 Design and analysis of dual-core dispersion compensating leaky fibre *12th Int. Symposium on Microwave and Optical Technology – ISMOT 2009 (New Delhi, Dec. 16 – 19) (Macmillan Publishers India Ltd.)* eds Sharma E K, Verma A K, Gupta R S, Gupta M, Saxena M, and Singh S pp 629-32.

[9]. Dussardier B, Trzesien S, Ude M, Rastogi V, Kumar A and Monnom G 2008 Large mode area leaky optical fibre fabricated by MCVD *Int. Conf. on Fiber Optics and Photonics – Photonics 2008 (New Delhi, India, Dec. 14-17) (New Delhi: Viva Books Pvt. Ltd.)* eds Pal B P and Sharma A p 88

[10]. Morishita K 1981 Numerical analysis of pulse broadening in graded index optical fibers *IEEE Trans. Microwave Theory Tech.* MTT-**29** 348–52

[11]. Thyagarajan K, Diggavi S, Taneja A, and Ghatak A K 1991 Simple numerical technique for the analysis of cylindrically symmetric refractive-index profile optical fibers *Appl. Opt.* **30** 3877–9

[12]. Chiang K S 1987 Radial effective index method for the analysis of optical fibers *Appl. Opt.* **26** 2969-73

[13]. Marcuse D 1982 Influence of curvature on the losses of doubly clad fibers *Appl. Opt.* **21** 4208-13

**Table 1**

MLC fibre parameters for LMA operation at 1.55 µm wavelength

| $n_1$ | $\Delta$ (%) | $a$ (µm) | $b$ (µm) | $d$ (µm) | $\Lambda$ (µm) |
|---------|------|------|--------|------|------|
| 1.44439 | 0.6 | 15 | 62.5 | 3 | 9 |

**Figure Captions**

**Figure 1.** (a) Schematic of transverse cross-section of an MLC leaky fibre. (b) The corresponding refractive index profile.

**Figure 2.** Comparison between FEM and TMM.

**Figure 3.** Bending losses for a circular-core MLC fibre for profile shape parameter $q = 1$.

**Figure 4.** Birefringence in an elliptical-core MLC Fibre. Dashed line represents the birefringence of two-layer conventional elliptical core fibre. The errorbars are due to finite accuracy for the chosen mesh.

**Figure 5.** Crosses and filled circles represent the variation of bending loss of $x$-polarized and $y$-polarized modes, respectively with bending radius for an elliptical-core MLC leaky fibre for $q = 1$. Filled squares show the variation of birefringence with bending radius for $q = 5$.

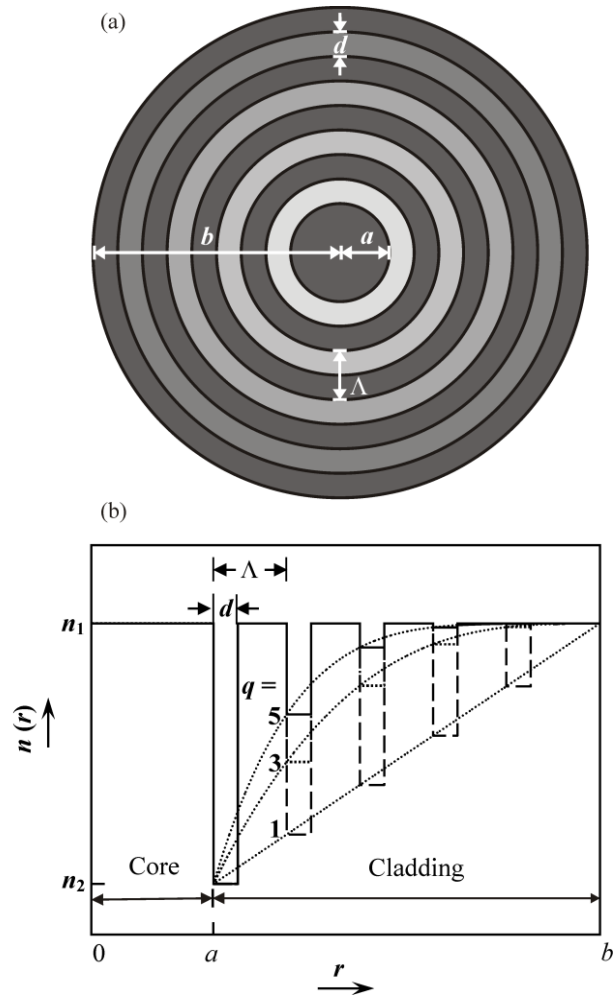

Figure 1
Labonté et al.

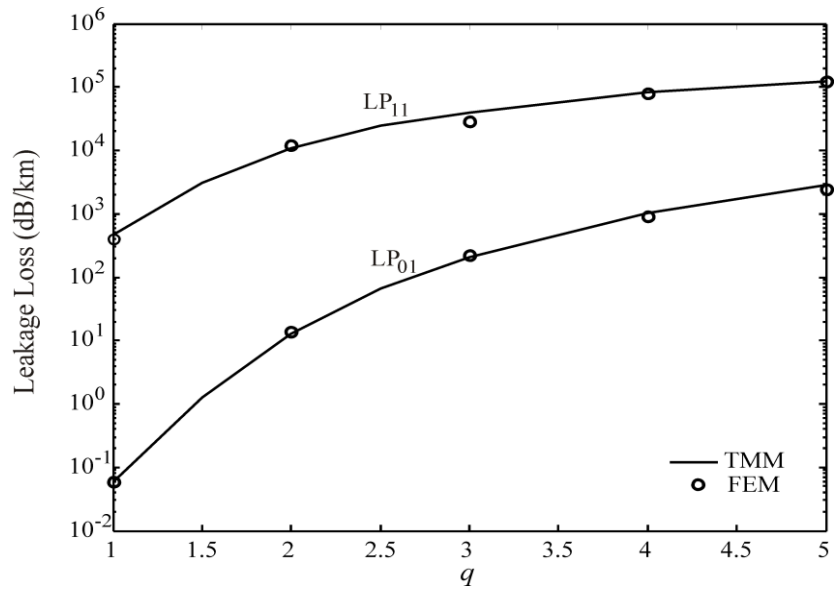

Figure 2
Labonté et al.

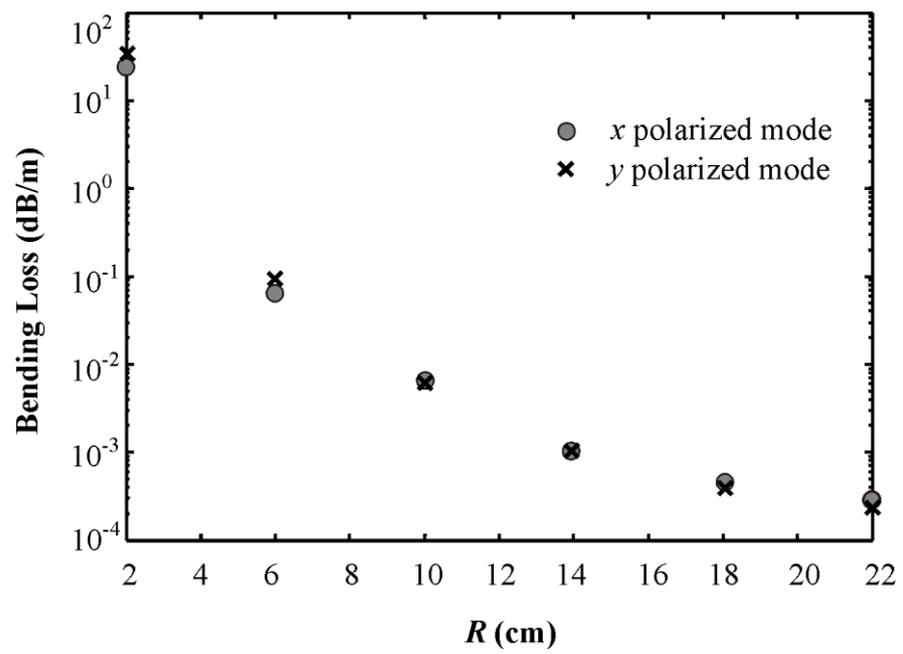

Figure 3
Labonté et al.

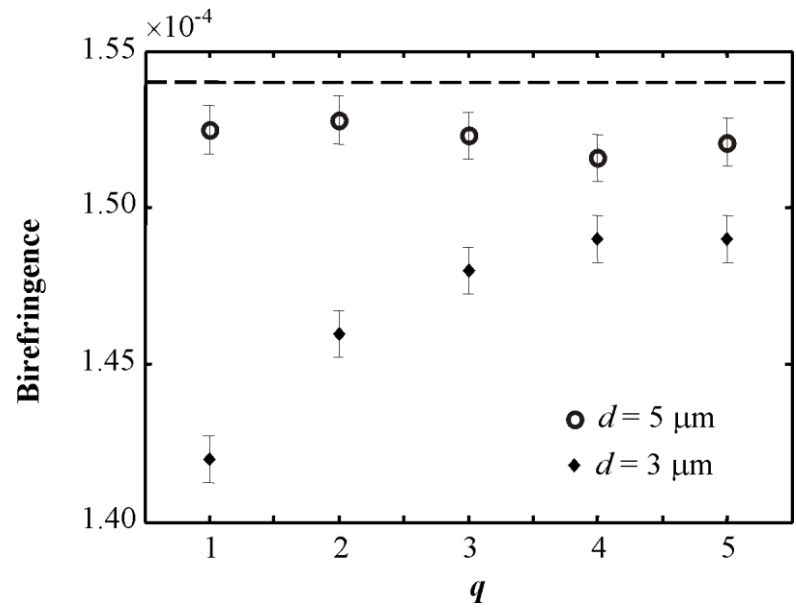

Figure 4
Labonté et al.

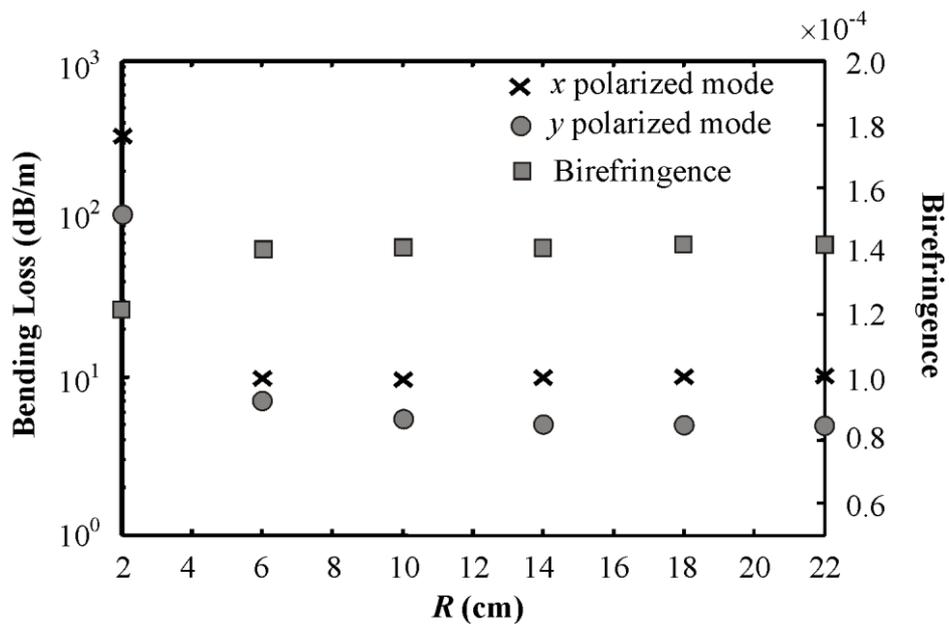

Figure 5
Labonté et al.